\documentclass[seceq]{ptptex}

\usepackage[dvips]{graphicx}
\usepackage[dvips]{color}

\usepackage[normalem]{ulem}  % \sout{old text} for strikeout

\renewcommand\sout{\bgroup \color{red} \ULdepth=-.5ex \ULset}

\newcommand{\Slash}[1]{\ooalign{\hfil/\hfil\crcr$#1$}}
\newcommand{\re}{\text{Re }}
\newcommand{\im}{\text{Im }}

\title{%        %You can use \\ for explicit line-break.
Structure of $\Lambda(1405)$ and threshold behavior 
of $\pi\Sigma$ scattering%
}

\author{%       %Use \scshape for the family name.
Yoichi \textsc{Ikeda}$^{1}$, %
Tetsuo \textsc{Hyodo}$^{2}$, %
Daisuke \textsc{Jido}$^{3}$, %
Hiroyuki \textsc{Kamano}$^{4}$, %
Toru \textsc{Sato}$^{5}$ and %
Koichi \textsc{Yazaki}$^{1,3}$%
}

\inst{%     %Affiliation, neglected when [addenda] or [errata].
$^{1}$Nishina Center for Accelerator-Based Science, Institute for Physical
and Chemical Research (RIKEN), Saitama 351-0198, Japan\\
$^{2}$Department of Physics, Tokyo Institute of Technology 
Meguro 152-8551, Japan\\
$^{3}$Yukawa Institute for Theoretical Physics, Kyoto University,
Sakyo, Kyoto 606-8502, Japan\\
$^{4}$Excited Baryon Analysis Center (EBAC), 
Thomas Jefferson National Accelerator Facility,
Newport News, Virginia 23606, USA\\
$^{5}$Department of Physics, Osaka University, Osaka 560-0043, Japan\\
}

\abst{%         %This abstract is neglected when [addenda] or [errata].
The scattering length and effective range of the $\pi\Sigma$ channel 
are studied in order to characterize the strangeness $S=-1$ meson-baryon scattering 
and the $\Lambda(1405)$ resonance. 
We examine various off-shell dependence of the amplitude 
in dynamical chiral models 
to evaluate the threshold quantities with the constraint 
at the $\bar{K}N$ threshold. 
We find that the $\pi\Sigma$ threshold parameters are important 
to the structure of the $\Lambda(1405)$ resonance 
and provide further constraints on the subthreshold extrapolation 
of the $\bar{K}N$ interaction.
}

\begin{document}

\maketitle

%%%%%%%%%%%%%%%%%%%%%%%%%%%%%%%%%%%%%%%%%%%%%%%%%%%%%%%%%%%%%%%%%%
\section{Introduction}
%%%%%%%%%%%%%%%%%%%%%%%%%%%%%%%%%%%%%%%%%%%%%%%%%%%%%%%%%%%%%%%%%%

The $\Lambda(1405)$ hyperon resonance has strangeness $S=-1$ and isospin $I=0$, and is located just below the threshold of $\bar KN$. 
Because $\Lambda(1405)$ is considered to be the quasi-bound state of the $\bar{K}N$ channel, the structure of $\Lambda(1405)$ is one of the very important issues of recent hadron physics, especially to understand $\bar K$-nucleon and $\bar K$-nucleus interactions. At the same time, $\Lambda(1405)$ is a resonance decaying to the $\pi \Sigma$ channel with $I=0$ by the strong interactions. Thus, for the nature of $\Lambda(1405)$, dynamics of both $\pi\Sigma$ and $\bar KN$ is important and gives essential contributions. 

The $\Lambda(1405)$ resonance
has been considered as a dynamically generated state in meson-baryon scattering for a long time before establishment of QCD, while the possibility of the $\Lambda(1405)$ with some three-quark components originated in quark-model viewpoint has been also discussed. The first dynamical investigation of $\Lambda(1405)$ in terms of meson and baryon degrees of freedom was performed by Dalitz 
%%%
and Tuan~\cite{Dalitz:1959dn}.
%%%
There $\pi \Sigma$ and $\bar KN$ scattering amplitudes were calculated in a coupled-channels approach and $\Lambda(1405)$ was obtained as a quasi-bound state of $\bar KN$. It was also confirmed in a modern point of view that $\Lambda(1405)$ can be predominantly described by meson-baryon components in a coupled-channels approach based on chiral dynamics~\cite{Hyodo:2008xr} (chiral unitary model), which successfully reproduced $\Lambda(1405)$ in meson-baryon dynamics~\cite{Kaiser:1995eg,Oset:1998it,Oller:2000fj,Lutz:2001yb,Oset:2001cn,Hyodo:2002pk,Jido:2003cb,Borasoy:2005ie,Borasoy,Cieply:2009ea}. A phenomenological approach also described $\Lambda(1405)$ as a quasi-bound state of $\bar KN$~\cite{AY02}, in which an effective interaction of $\bar KN$ was derived so as to reproduce the $\bar KN$ scattering length and the mass and width of $\Lambda(1405)$ as 1405 MeV and 40 MeV, as suggested by the Particle Data Group (PDG).\cite{Nakamura:2010zzi} Concerning the amplitude above the threshold, this effective potential is essentially consistent with the one extracted from the chiral unitary approach~\cite{Hyodo:2007jq}, although the phenomenological model provides quantitatively stronger $\bar KN$ interaction than the chiral potential in the region far below the $\bar{K}N$ threshold. 
Thus, there are uncertainties about the theoretical extrapolation of the $\bar K N$ interaction below the $\bar K N$ threshold.

Another recent finding is that $\Lambda(1405)$ is composed by two resonance states appearing around $\Lambda(1405)$ energies which have different coupling nature to the $\pi\Sigma$ and $\bar KN$ channels~\cite{Oller:2000fj,Jido:2003cb,fink, jido3}. According to Ref.~\citen{Jido:2003cb},  one state which dominantly couples to the $\bar KN$ state is located at 1420 MeV instead of nominal 1405 MeV with a narrower width, while the other state couples strongly to the $\pi\Sigma$ channel and appears around 1390 MeV with a wider width. Because of the dominant coupling to the $\bar KN$, the higher pole is more relevant to the $\bar K$-nucleus interaction~\cite{Hyodo:2007jq}. Thus, it is extremely important to reveal the $\Lambda(1405)$ resonance position in the $\bar KN$ channel. This double pole structure of $\Lambda(1405)$ also leads to the feature that the $\Lambda(1405)$ resonance spectra depend on its production mechanism~\cite{Jido:2003cb,Hyodo:2004vt,magas,Geng:2007hz,Jido:2009jf}. For observation of $\Lambda(1405)$ spectra below the $\bar KN$ threshold, an old bubble-chamber experiment in $K^{-} d \to \pi^{+}\Sigma^{-} n$ 
gave a hint of
the $\Lambda(1405)$ resonance position at 1420 MeV~\cite{Braun:1977wd}.
A recent theoretical analysis based on the chiral unitary approach~\cite{Jido:2009jf} confirmed that the $\Lambda(1405)$ production is initiated by the $\bar KN$ channel in the $K^{-}d \to \pi\Sigma n$, and was able to reproduce the $\Lambda(1405)$ production cross section and  the shape of the experimental observed spectrum.
Further detailed experimental information of the $\pi\Sigma$ spectrum will be obtained in ongoing/forthcoming experiments at Jefferson Laboratory~\cite{Moriya:2009mx}, GSI~\cite{Siebenson:2010hh}, and J-PARC.~\cite{Noumi:JPARC}

Recent theoretical investigations based on coupled-channels approaches in chiral dynamics suggested the double pole structure of $\Lambda(1405)$. Nevertheless, the position of the lower pole is dependent on the details of the model parameters, as discussed in Ref.~\citen{Hyodo:2007jq}. Moreover, Ref.~\citen{Ikeda:2010tk} shows that the use of the approximate energy independent potential also changes the pole position drastically. Indeed, since the pole position of the higher state is strongly constrained by observed $K^{-} p$ scattering and $K^{-}$ hydrogen experiments, the theories provide very similar pole positions around 1420 MeV with narrow widths within the framework of several chiral unitary models. However, due to lack of $\pi\Sigma \to \pi\Sigma$ experimental data, the properties of the lower state are less controlled in theoretical studies. Hence the pole position of the lower state has model dependence even in similar models constructed by the same concept. The $\pi\Sigma$ dynamics, including the nature of the lower $\Lambda(1405)$ pole, may be important in kaonic nuclear few-body systems. A possible quasibound state of $\bar KNN$ was proposed in 60's~\cite{nogami}, and recent theoretical investigations concluded that the $\bar KNN$ system has a quasibound state with a large width~\cite{AY02,Ikeda:2007nz,shevchenko07,dote08,Wycech:2008wf,ikeda09,Ikeda:2010tk}, and the theoretical predictions of the energy and width scatter over a wide range.
Although part of the ambiguity in the model prediction has been attributed to the treatment of the three-body dynamics~\cite{ikeda09} and the energy dependence of the interaction~\cite{Ikeda:2010tk}, it is obvious that the present experimental database does not completely constrain the $\bar{K}N$-$\pi\Sigma$ interaction. Since the $\pi\Sigma$ threshold is located at 100 MeV below the $\bar KN$ threshold, if $\bar K$ is bound in few-body systems with a large binding energy, the coupling to the $\pi \Sigma$ channel is essential to understand the properties of the quasibound $\bar KNN$ state.

In this paper, we discuss threshold behavior of the $\pi\Sigma$ scattering\footnote{A preliminary discussion was given in Ref.~\citen{Jido:2010ag}.}, emphasizing its importance in a context of the $\Lambda(1405)$ resonance in chiral dynamics.  For the physics of $\Lambda(1405)$, it is certainly necessary to have theoretical descriptions of $\pi\Sigma$ and $\bar KN$ dynamics applicable for the energy region between their thresholds, which is as wide as 100 MeV. The nature and position of the pole singularity (bound state, virtual state, or resonance) around the $\pi\Sigma$ threshold\footnote{Here we investigate the coupled-channels scattering amplitudes below the $\bar KN$ threshold. Thus, we search the poles in the $\bar KN$ physical sheet of the complex energy plane and classify the poles in terms of the Riemann sheets for the $\pi\Sigma$ channel.}
is an important issue. 
Because chiral symmetry indicates that the $s$-wave $\pi\Sigma$ diagonal interaction at low energy with $I=0$ is strongly attractive, a pole singularity may be produced at least somewhere, probably close to the threshold, in the complex energy plane. 
It is known that the leading order calculations of the pion-nucleon and pion-pion scattering lengths in the chiral perturbation theory work well within 20\% accuracy. Thus, it is an interesting question whether it is also the case for the scattering length for 
the $\pi\Sigma$ channel with $I=0$.
To clarify the nature of the $\pi\Sigma$ dynamics, we calculate the scattering length $a_{\pi\Sigma}$ and the effective range $r_e$ in various models of the $\pi\Sigma$ and $\bar KN$ coupled channels in which different ways of solving Lippmann-Schwinger equation are applied in the isospin limit. Finally, we conclude that the $\pi\Sigma$ scattering length and effective range give us new and essential information of the $\pi\Sigma$-$\bar KN$ dynamics between their thresholds and the $\Lambda(1405)$ resonance. We discuss the general relationship between the scattering length and the effective range with the pole singularity around the threshold in Appendix.

%%%%%%%%%%%%%%%%%%%%%%%%%%%%%%%%%%%%%%%%%%%%%%%%%%%%%%%%%%%%%%%%%%
\section{Two channel models of $\bar{K}N$-$\pi \Sigma$ scattering}\label{sec:2channel}
%%%%%%%%%%%%%%%%%%%%%%%%%%%%%%%%%%%%%%%%%%%%%%%%%%%%%%%%%%%%%%%%%%

We discuss the threshold behavior of the $\pi \Sigma$ amplitude 
with $I=0$ in connection with the $\Lambda(1405)$ resonance 
appearing below the $\bar KN$ threshold. 
To describe the $\pi\Sigma$ scattering amplitude and the $\Lambda(1405)$ resonance, 
it is essential to treat both $\pi \Sigma$ and $\bar KN$ channels simultaneously 
in a coupled-channels approach. 
Here we consider simple models of coupled $\bar KN$ and $\pi\Sigma$ channels with $I=0$ based on chiral effective theory. The interaction kernel is given by the Weinberg-Tomozawa interaction and we solve coupled-channel Lippmann-Schwinger equation with this interaction kernel. In our model, there are two adjustable parameters which come from the regularization of the Lippmann-Schwinger equation. These parameters are determined by the $\bar KN$ scattering length with $I=0$.
We shall also compare different theoretical schemes for solving Lippmann-Schwinger equation for the $\bar KN$-$\pi\Sigma$ coupled system.

\subsection{Framework of the models}
\label{sec:form}

Let us start with Lippmann-Schwinger equation for the $t$-operator 
of the $\pi\Sigma$ and $\bar KN$ channels with $I=0$:
\begin{equation}
    \hat{t}(W)  = \hat{v} +  \hat{v} \hat{g}(W) \hat{t}(W) \ ,  \label{eq:LSeq}
\end{equation}
where $W$ is the scattering energy in the center-of-mass frame, 
$\hat{v}$ is the interaction kernel, 
and $\hat{g}$ is the free two-body Green operator. 
We solve the Lippmann-Schwinger equation in the center-of-mass system. 
Inserting complete sets of the meson-baryon states, 
we obtain the Lippmann-Schwinger equation for the $t$-matrix element:
\begin{eqnarray}
   \lefteqn{
   \langle k_{i} | \hat{t}(W) | k_{j} \rangle  } && \nonumber \\ &=&
    \langle k_{i} | \hat{v} | k_{j} \rangle
      +\sum_{m,n} \int \frac{d^{3} q_m}{ (2\pi)^{3}  \sqrt{{\cal N}_m}} 
      \frac{d^{3}q_n}{(2\pi)^{3} {\sqrt{{\cal N}_n}}}
         \langle k_{i} | \hat{v} | q_m \rangle 
         \langle q_m | \hat{g}(W) | q_n \rangle 
      \langle q_n | \hat{t}(W) | k_{j} \rangle  , \ \ \ \ \ \label{eq:LSeqTele}
\end{eqnarray}
where $k_{i}$ and $k_{j}$ are final and initial meson (baryon) momenta 
with the channel indices $i,j$, respectively.
In general, we need off-shell interaction kernel and 
$t$-matrix element to solve Eq.~(\ref{eq:LSeqTele}). 
In this equation, 
the two-body meson-baryon state is normalized as
$\langle k_i | k_j \rangle = \delta_{ij} \sqrt{ {\cal N}_{i} {\cal N}_{j}}
(2\pi)^{3} \delta^{(3)}(\vec k_i - \vec k_j)$ 
where ${\cal N}_i =2 \omega_i E_i/M_i$ with the baryon mass $M_i$, 
the baryon energy $E_i(k_i)=\sqrt{k_i^{2}+M_i^{2}}$, 
the meson energy  $\omega_i(k_i)= \sqrt {k_i^{2}+m_i^{2}}$ and the meson mass $m_i$. 
The on-shell meson (baryon) momentum 
$k_i$
is given by
\begin{equation} 
    k_i= \frac{\sqrt{(W^{2}-(M_i+m_i)^{2})(W^{2}-(M_i-m_i)^{2})}}{2W} \ .
\end{equation}
 
We fix the interaction kernel $\hat{v}$ based on the chiral effective theory. 
At the leading order of the chiral expansion for the $s$-wave scattering, 
we take so-called Weinberg-Tomozawa interaction given by 
\begin{equation}
    \langle k_{i} | \hat{v} | k_{j} \rangle  = - \frac{C_{ij}}{4 f^{2}_{\pi}} \bar u_{i} 
    (\Slash{k_{i}}+ \Slash{k_{j}}) u_{j} \ , \label{eq:WTterm}
\end{equation}
where $i$ and $j$ are channel indices of $\pi\Sigma$ ($i,j=1$) 
and $\bar KN$ ($i,j=2$), $\vec k_{i}$ is the meson momentum for channel $i$
%%%
and $k^{0}_{i}$ is fixed to the on-shell energy $k^{0}_{i}=\omega_{i}(k_i)$
%%%
in the center-of-mass system, 
$f_{\pi}$ is the meson decay constant, 
$u_{i}$ is the Dirac spinor for the baryon and 
$C_{ij}$ is the coupling strength which is determined by the flavor SU(3) symmetry as
\begin{equation}
   C = \left(
   \begin{array}{cc}
       4 & -\sqrt{\frac{3}{2}} \\
       -\sqrt{\frac{3}{2}} & 3
   \end{array}
   \right).  \label{eq:Cij}
\end{equation}
Note that the 
%%%
meson-energy-momentum
%%%
dependence of the interaction~\eqref{eq:WTterm} 
is the consequence of the Nambu-Goldstone theorem, 
which is manifested as the derivative coupling in the effective chiral Lagrangian. 
Although the flavor SU(3) symmetry requires the full coupled channels 
with $\eta\Lambda$ and $K\Xi$, 
it is found that the effect of these channels are small to the scattering amplitude 
in the energy region of interest~\cite{Hyodo:2007jq}.

With the Weinberg-Tomozawa interaction, 
which is a contact interaction in the coordinate space, 
we need to regularize the momentum integration appearing 
in the Lippmann-Schwinger equation~(\ref{eq:LSeqTele}).  
As we shall explain 
%%%
in \S \ref{sec:modelA} and \ref{sec:modelB},
%%%
we will use two
%%%
kinds of
%%%
renormalization schemes. 
The regularization is done in each channel, 
so that we have two parameters in the models in the form of 
cut-offs or subtraction constants for the $\pi\Sigma$ and $\bar KN$ channels. 
These parameters should be chosen so as to reproduce available experimental data. 

In this work, 
we would like to investigate how the $\pi\Sigma$ threshold behavior is constrained 
by the $\bar KN$ scattering length. 
For reference, 
we use a value of the $\bar KN$ scattering length with $I=0$, 
$a_{\bar KN} = -1.70+0.68i$ fm, 
determined by 
%%% 
Martin in
%%%
Ref.~\citen{Martin:1980qe}. 
This strategy 
%%%
was also taken in constructing
%%%
 ``Model (a)'' in Ref.~\citen{Ikeda:2007nz}. 
Since the scattering length is a complex value, 
this is enough to determine the two cut-off parameters 
for the $\pi\Sigma$ and $\bar KN$ channels. 
This value of the $\bar KN$ scattering length leads to the $\Lambda(1405)$ pole position 
around 1420 MeV~\cite{Ikeda:2007nz}. 

In the following subsections, 
we will explain further details of our models, 
which are obtained by different treatments of the regularization 
of the integral and the interaction kernel. 
These theoretical choices stem from ways of off-shell treatment 
of the scattering amplitudes. 
In our models, 
we assume isospin symmetry 
%%%
and use
%%%
isospin averaged masses, \label{mass}
$m_{\pi}=138$ MeV, $M_{\Sigma}=1193$ MeV, $M_{\bar K}=496$ MeV 
and $M_{N}=939$ MeV, 
and the meson decay constant $f_{\pi}$ is fixed to be $f_{\pi}=92.4$ MeV. 
Note that one should utilize the physical masses for a realistic calculation 
of the threshold observables, 
because the isospin breaking effects may be quantitatively large 
at the threshold.~\cite{Meissner:2004jr} 
In addition, for direct comparison to experimental observation, Coulomb effects 
should also
%%%
be taken into account as done in Refs.~\citen{Dalitz:1959dn,Borasoy:2005ie} for the analyses of $\bar KN$ scattering.
However, 
the present work focuses on a qualitative examination
of the nature of the $\pi\Sigma$ channel, and precise evaluations of the 
$\pi\Sigma$ scattering length and effective range are out of the scope of the 
present paper. 
Thus
we believe that the
calculation in the isospin basis using the isospin averaged masses without Coulomb corrections
%%%
is sufficient for the present purpose.
%%%

After obtaining the $s$-wave scattering amplitudes 
in our dynamical models, 
the scattering length $a$ and the effective range $r_e$ in channel $i$ are extracted by
\begin{align}
    a =& f_i(k_i)|_{k_i\to 0} \ , \\
    r_e =&
    \left.\frac{d^2}{dk_i^2}\left(\frac{1}{f_i(k_i)}\right)
    \right|_{k_i\to 0} \ ,
\end{align}
where the $s$-wave scattering amplitude $f_i(k_i)$ is given by
\begin{equation}
    f_i(k_{i})   =  -\frac{M_i}{4\pi W} \langle k_i | \hat{t}(W) | k_i \rangle  \ .
\end{equation}
The definition of $a$ and $r_{e}$ is given with the phase shift $\delta_{i}$ by 
\begin{equation}
    k_{i} \cot \delta_{i} = \frac{1}{a} + r_e\frac{k^2_{i}}{2}
    +\cdots .
\end{equation}
%%%

\subsection{Model A}
\label{sec:modelA}

Model A is based on a chiral unitary approach given in Ref.~\citen{Hyodo:2002pk}. 
In this model, the interaction kernel for $s$-wave is given by
\begin{equation}
   V_{ij}(W) \equiv \langle k_i | \hat{v} | k_j \rangle =
    - \frac{C_{i j} }{4 f_{\pi}^2}(2W - M_{i}-M_{j})
    \sqrt{\frac{M_{i}+E_{i}}{2M_{i}}}
    \sqrt{\frac{M_{j}+E_{j}}{2M_{j}}} \ ,
\end{equation}
with $C_{ij}$ given in Eq.~(\ref{eq:Cij}). 
According to the $N/D$ method with a simplification of neglecting 
the left-hand cut by $N=1$, 
which was developed in Ref.~\citen{Oller:2000fj}, 
we can use the on-shell values of the interaction kernel 
in the Lippmann-Schwinger equation (\ref{eq:LSeqTele}). 
This corresponds to implementing the elastic unitarity. 
Then the matrix elements of $\hat{v}$ and $\hat{t}$ 
inside 
the $\vec q_n$ and $\vec q_m$ integrals in Eq.~(\ref{eq:LSeqTele}) 
can be factorized out from the integrals, and one of the momentum integral can be trivially performed because the Green function is diagonal. 
Therefore, we only perform the integral of the free Green function. 
Applying dimensional regularization, we obtain
%%%
the finite part of the Green function as
%%%
\begin{eqnarray}
  \lefteqn{ 
  G_{i}(W)  \equiv
  \int \frac{d^3 q_i d^3 q_j} {(2\pi)^{6} \sqrt{ {\cal N}_i{\cal N}_j}}
  \langle q_i | \hat{g}(W) | q_j \rangle 
} && 
 \nonumber \\
  &=&  i \,  \int \frac{d^4 q_i}{(2 \pi)^4} \,
\frac{2 M_i}{q_i^2 - M_i^2 + i \epsilon} \, \frac{1}{(P-q_i)^2 - m_i^2 + i
\epsilon} \nonumber \\
 &=& \frac{2 M_i}{16 \pi^2} \left\{ 
{d_i(\mu)} +\ln \frac{M_i^2}{\mu^2} \right. 
  + \frac{m_i^2-M_i^2 + W^{2}}{2W^{2}} \ln \frac{m_i^2}{M_i^2} 
\nonumber \\ &  +&
\frac{k_{i}}{W}
\left[ \ln(W^{2}-(M_i^2-m_i^2)+2 k_i W) \right. 
+ \ln(W^{2}+(M_i^2-m_i^2)+2 k_i W) 
 \nonumber \\
 &-& \left. \left.  \ln(-W^{2}+(M_i^2-m_i^2)+2 k_i W) 
   - \ln(-W^{2}-(M_i^2-m_i^2)+2k_i W) \right]
\right\} ,
\label{eq:gpropdr}
\end{eqnarray}
where 
$P_{\mu}=(W,0)$ is the total energy momentum in the CM frame,
%%%
$\mu$ is the scale of dimensional regularization,
%%%
which is to be fixed at  $\mu=630$ MeV here, 
and $d_i(\mu)$ is the remaining regularized finite constant term.
We call $d_i(\mu)$ subtraction constant. The subtraction constants,
$d_{\bar KN}$ and $d_{\pi \Sigma}$, are adjustable parameters in this model 
and to be fixed by experimental data or theoretical requirement. 
Here we determine the subtraction constants by the $\bar KN$ scattering length.
%%%
%%%

After having done the regularization of the loop integral, 
we can solve the Lippmann-Schwinger equation in an algebraic way:
\begin{equation}
   \langle k_i | \hat{t}(W) | k_j \rangle = 
   \sum_n 
   \Bigl[ \frac{1}{1-V(W)G(W)} \Bigr]_{in} V_{nj}(W) .
\end{equation}

%%%%%%%%%%%%%%%%%%%%%%%%%%%%%%%%%%%%%%%%%%%%%%%%%%%%%%%%%%%%%%%%%%%%%
\subsection{Model B}
\label{sec:modelB}
%%%%%%%%%%%%%%%%%%%%%%%%%%%%%%%%%%%%%%%%%%%%%%%%%%%%%%%%%%%%%%%%%%%%%
%%%
Here we explain another potential model (Model B) for the $\pi\Sigma$ scattering 
which is developed in Ref.~\citen{Ikeda:2007nz}. 
In this model, 
the interaction kernel on the mass-shell 
is given by
\footnote{Since we are using a different normalization 
of the $t$-matrix element from Ref.~\citen{Ikeda:2007nz}, 
we have some trivial difference in the factor of Eq.~(\ref{eq:WTEdep}) 
than the original form in Ref.~\citen{Ikeda:2007nz}.}
\begin{equation}
    \langle k_i | \hat{v} | k_j \rangle = 
    F_i(k_i) \lambda_{ij}(W) F_j(k_j)
    = - \frac{C_{ij}}{4f_{\pi}^{2}}(2W-M_{i}-M_{j})
    F_i(k_i) F_j(k_j) \ ,  \label{eq:WTEdep}
\end{equation}
with
\begin{equation}
   F_{i}(k_{i}) = \frac{\Lambda_{i}^{4}}{(k_{i}^{2}+\Lambda_{i}^{2})^{2}} \ ,
\end{equation}
where $C_{ij}$ is given in Eq.~(\ref{eq:Cij}), 
and 
$F_i(k_i)$ is a separable form factor of dipole type.
Since we introduce the form factors in Eq~(\ref{eq:WTEdep}), 
the on-shell kernel has different form used in Model A.
It is, however, noticed that $\lambda_{ij}(W)$ has the same energy dependence
as the interaction kernel used in Model A 
except the relativistic correction factor.

For the off-shell behavior of the interaction kernel, 
using the same form of the separable form factor
as in Eq.~(\ref{eq:WTEdep}) with off-shell momenta, we define 
\begin{equation}
    \langle q_{i} | \hat{v}(W) | q_{j} \rangle 
    = F_{i}(q_{i}) \lambda_{ij}(W)  F_{j}(q_{j}) \ .  \label{eq:off}
\end{equation}
The cut-off parameter appearing in each channel will be fixed by the inputs later. 
We call the model with this energy dependent interaction by ``Model B E-dep''.

We also consider an energy independent interaction given by
%%%
taking the 0th components of $k^{\mu}_{i}$ and $k^{\mu}_{j}$ 
of Eq.~(\ref{eq:WTterm}) and the static approximation:
%%%
\begin{eqnarray}
    \langle k_i | \hat{v} | k_j \rangle &=&
    - \frac{C_{ij}}{4f_\pi^2}(\omega_i(k_i) + \omega_j(k_j)) F_i(k_i)F_j(k_j) \nonumber \\
               & \simeq & - \frac{C_{ij}}{4f_\pi^2}(m_i + m_j) F_i(k_i)F_j(k_j) \ .
     \label{eq:WTEindep}
\end{eqnarray}
%%%
These kinds of the energy independent potential are often used in few-body 
calculations and are favorable for non-relativistic quantum calculations.
%%%
The off-shell behavior is defined in the same way as Eq.~(\ref{eq:off}). 
We call the model with this energy-independent interaction by ``Model B E-indep''. 
This energy-independent potential may contradict
the energy expansion scheme in chiral effective theory at energies far from the threshold. 
Nevertheless, it is interesting to compare the results with this potential and others, 
in order to see the importance of energy dependence of the potential. 

Since we introduce the form factor in the interaction kernel, the integration is regularized:
\begin{eqnarray}
  \lefteqn{
  \sum_{m,n}
  \int \frac{d^3 q_m d^3 q_n}{ (2\pi)^{6} \sqrt{{\cal N}_m {\cal N}_n}}
  \langle k_i | \hat{v} | q_m \rangle 
  \langle q_m | \hat{g}(W) | q_n \rangle 
  \langle q_n | \hat{t}(W) | k_j \rangle
  } && \nonumber \\
  &=& F_{i}(k_i) F_{j}(k_j)
  \sum_{n=1}^{2}  \lambda_{in}(W) \tau_{nj}(W) 
  \int \frac{q^{2}_{n} d {q_{n}}}{2\pi^{2}}  \frac{1}{ 2\omega_{n}} 
  \frac{F_{n}(q_{n}) F_{n}(q_{n})}{W-E_{n}-\omega_{n} + i\epsilon} \ ,
\label{pot-LS}
\end{eqnarray}
with
\begin{equation*}
\langle k_i | \hat{t}(W) | k_j \rangle = F_i(k_i) \tau_{ij}(W) F_j(k_j)\ ,
\end{equation*}
with $E_{n}=\sqrt{q_{n}^{2} + M_{n}^{2}}$ and $\omega_n=\sqrt{q_{n}^{2} + m_{n}^{2}}$.
In Eq. (\ref{pot-LS}),
we use the fact that the $t$-matrix elements can be factorized in separable forms
when the separable interactions are taken into account.
The $t$-matrix element is obtained again in an algebraic way as
\begin{equation}
   \langle k_{i} | \hat{t}(W) | k_{j} \rangle =
   F_{i}(k_i) F_{j}(k_j) 
   \sum_n \Bigl[ \frac{1}{1-\lambda(W) G(W) } \Bigr]_{in} 
   \lambda_{nj}(W)  \ ,
\end{equation}
where
\begin{equation}
  G_{n}(W)  \equiv 
    \int \frac{q^{2}_{n} d {q_{n}}}{2\pi^{2}}  \frac{1}{ 2\omega_{n}} 
  \frac{F_{n}(q_n) F_{n}(q_{n})}{W-E_{n}-\omega_{n} + i\epsilon}  \ .
\end{equation}

%%%%%%%%%%%%%%%%%%%%%%%%%%%%%%%%%%%%%%%%%%%%%%%%%%%%%%%%%%%%%%%%%%
\section{Threshold parameters in $\pi\Sigma$ channel and $\Lambda(1405)$}
%%%%%%%%%%%%%%%%%%%%%%%%%%%%%%%%%%%%%%%%%%%%%%%%%%%%%%%%%%%%%%%%%%

In this section, 
we show the results of the $\pi\Sigma$ scattering length and effective range 
obtained by our two-channel models. 
First of all, we fix the parameters appearing in each model 
so as to reproduce the $\bar KN$ scattering length $a_{\bar KN} = -1.70+0.68i$ fm
%%%
\cite{Martin:1980qe}.
%%% 
To characterize the subthreshold behavior of the amplitude, 
we also calculate the pole positions of the $\Lambda(1405)$ resonance 
in the complex energy plane.
%%%
The poles are searched in the $\bar KN$ physical Riemann sheet and
classified in terms of the two $\pi\Sigma$ sheets; bound states in the first Riemann
sheet, while virtual states and resonances are in the second Riemann sheet.
%%%

Let us first show the result of Model A 
with dimensional regularization of the loop function. 
We find two solutions to reproduce the $\bar{K}N$ scattering length ($a_{\bar KN}=-1.70+0.68i$ fm); 
we shall call these two solutions ``Model A1'' and ``A2'':
\begin{align}
    d_{\pi\Sigma}
    &=-1.67,&
    d_{\bar{K}N}
    &=-1.79 , &&
     \text{(Model A1)} ,
    \label{eq:Martin1} \\
    d_{\pi\Sigma}
    &=-2.85,  &
    d_{\bar{K}N}
    &=-2.05 , && \text{(Model A2)} .
    \label{eq:Martin2}
\end{align}
Note that the natural value of the subtraction constant~\cite{Hyodo:2008xr} is 
$d_{\bar{K}N}=-1.95$ and $d_{\pi\Sigma}=-1.60$ at this scale, 
so that  Model A1 is more natural from the theoretical point of view.

With the subtraction constant~\eqref{eq:Martin1}, 
the $\pi\Sigma$ scattering length and effective range are obtained as
%\begin{align}
%    &a_{\pi\Sigma}=0.934 \text{ fm} , && r_{e}= 5.02 \text{ fm},
%    && \text{(Model A1)}  . 
%    \nonumber \\
%\end{align}
\begin{equation}
    a_{\pi\Sigma}=0.934 \text{ fm} , \quad r_{e}= -5.02 \text{ fm},
    \quad \text{(Model A1)}  .
    \nonumber 
\end{equation}
%\intertext{
The pole singularities in the scattering amplitudes are found at
%}
%\begin{align}
%    &z=1422-16 i \text{ MeV}, &
%    &z=1375-72 i \text{ MeV}, && \text{(Model A1)} .
%    \nonumber
%\end{align}
\begin{equation}
    z=1422-16 i \text{ MeV}, \quad
    z=1375-72 i \text{ MeV}, \quad \text{(Model A1)} .
    \nonumber
\end{equation}
These poles correspond to resonance states 
located between $\bar{K}N$ and $\pi\Sigma$ thresholds. 
The pole positions obtained from Model A1
are similar 
to the ones obtained in standard parameterizations of chiral unitary models; 
the $\bar{K}N$ bound state and the $\pi\Sigma$ resonance.\cite{Jido:2003cb,Hyodo:2007jq}

The other solution (Model A2) in Eq.~(\ref{eq:Martin2}) provides 
the $\pi\Sigma$ scattering length and effective range as
\begin{equation}
    a_{\pi\Sigma}=-2.30 \text{ fm}, \quad r_{e} = -5.89 \text{ fm},
    \quad \text{(Model A2)}    ,
    \nonumber 
\end{equation}
and we find the pole singularities at
\begin{equation}
    z=1425-11 i \text{ MeV}, \quad
    z=1321 \text{ MeV} \ (\text{bound state}),
    \quad \text{(Model A2)}.  
    \nonumber
\end{equation}
In this case, 
one pole appears at $ z=1425-11 i $ MeV which may be interpreted as 
the $\Lambda(1405)$ resonance, 
while the other pole is found in the first Riemann sheet 
of the $\pi\Sigma$ channel below the threshold, 
corresponding to a bound state of $\pi\Sigma$. 
Following the discussion in Ref.~\citen{Hyodo:2008xr}, 
the large negative value of the subtraction constant 
is equivalent to the enhancement of the interaction strength. 
Thus, this parameter set provides the $\pi\Sigma$ interaction 
stronger than Model A1 and the scattering length in the $\pi\Sigma$ channel becomes negative. 
A general discussion of the relation of the scattering length and pole position
of scattering amplitude is given in Appendix. 

In Model B, we utilizes the separable potential with form factors 
and have two options of the energy dependence of 
the interaction kernel \eqref{eq:WTEdep} and \eqref{eq:WTEindep}. 
The $\bar KN$ scattering length $a_{\bar KN}=-1.70 + i0.68$ fm is reproduced by the following cut-off parameters:
\begin{align}
    \Lambda_{\pi\Sigma}& = 1005 \text{ MeV}, &
    \Lambda_{\bar KN}& = 1188 \text{ MeV}, && \text{ (Model B E-dep)}, \\
    \Lambda_{\pi\Sigma}& = 1465 \text{ MeV},&
    \Lambda_{\bar KN}& = 1089 \text{ MeV}, && \text{ (Model B E-indep)}.
\end{align}
With these parameters, the $\pi\Sigma$ scattering length and effective range are obtained as
\begin{align}
   a_{\pi\Sigma} &=  1.44 \text{ fm}, & r_{e}&=-3.96 \text{ fm},
   &&\text{ (Model B E-dep)}, \\
   a_{\pi\Sigma} &=  5.50 \text{ fm}, & r_{e}& =-0.458\text{ fm}, 
   &&\text{ (Model B E-indep)}.
\end{align}
and the pole positions are
\begin{align}
   z&=1422-22i \text{ MeV}, &z&=1349-54i \text{ MeV},
    &&\text{ (Model B E-dep)}, \\
   z&=1423-29i \text{ MeV},  &z&=1325 \text{ MeV}\ (\text{virtual state}),
   &&\text{ (Model B E-indep)}.
\end{align}
The result of Model B E-dep is very similar to that of Model A1, although different regularization schemes are applied. For the energy independent interaction case, one of the poles is obtained as a virtual state located in the second Riemann sheet of the $\pi \Sigma $ channel below threshold.
Thus, the scattering length is to be a large positive value. This is also understood by the comparison of the cutoff parameters in the model. Since the large cutoff parameter effectively enhances the interaction strength, the $\pi \Sigma$ attraction in Model B E-indep is stronger than that of Model B E-dep, leading to the formation of a virtual state below the threshold. It should be noted that Model B E-indep also develops two poles as shown in Ref.~\citen{Ikeda:2010tk}, although one of them is a virtual state.

\begin{figure}[tbp]
    \centering
    \includegraphics[width=0.8\textwidth,clip]{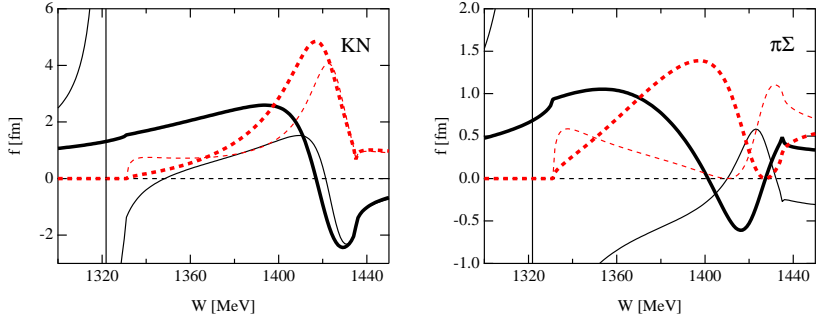}
    \caption{\label{fig:amplitudeA}
    Scattering amplitudes in $\bar{K}N$ (left panel) and $\pi\Sigma$ (right panel)
    channels.
    The thick (thin) curves represent the results from model A1 (A2), 
    and real and imaginary parts are plotted by solid and dashed curves in
    both panels.}
\end{figure}

We would like to emphasize that all the parametrizations are constrained by the $\bar{K}N$ scattering length very well, but $\pi\Sigma$ scattering lengths and pole positions in these parametrizations are quite different. This means that only with the $\bar KN$ threshold behavior, the $\pi\Sigma$ and $\bar KN$ coupled-channels amplitudes cannot be determined, even though the interaction kernels are fixed by theoretical consideration. This is clearly seen by plotting the scattering amplitude $f$ as functions of $W$ (Fig.~\ref{fig:amplitudeA}). In the left panel, the $\bar{K}N$ amplitudes around threshold coincide with each other since we use the $\bar{K}N$ scattering length to fix the model parameters. The agreement of the two models holds down to $W\sim 1420$ MeV, but below that energy, the extrapolation 
% is out of control 
cannot be controlled 
by the $\bar{K}N$ scattering length. The difference is significant in the right panel where the amplitude of the $\pi \Sigma$ channel is plotted. Of course, we have the experimental data of the $\pi\Sigma$ mass spectrum and the threshold branching ratios $K^{-}p \to \pi \Sigma$ and $\pi \Lambda$ observed in decay of kaonic hydrogen~\cite{Tovee:1971ga,Nowak:1978au}, which should reduce the difference between models. It is nevertheless important to keep in mind that the $\bar{K}N$ scattering length does not fully constrain the structure of the $\Lambda(1405)$ resonance and the $\bar{K}N$ amplitude far below the threshold.

%--figure---------------------------------
%--figure---------------------------------

%%%%%%%%%%%%%%%%%%%%%%%%%%%%%%%%%%%%%%%%%%%%%%%%%%%%%%%%%%%%%%%%%%
\section{Discussion}\label{sec:discussion}
%%%%%%%%%%%%%%%%%%%%%%%%%%%%%%%%%%%%%%%%%%%%%%%%%%%%%%%%%%%%%%%%%%

We have 
investigated
%%%
the scattering length and effective range 
of the $\pi\Sigma$ channel with $I=0$ using the two-channel models 
of the $\pi\Sigma$ and $\bar KN$ coupled systems. 
We have considered two kinds of models with different ways 
to solve the Lippmann-Schwinger equation. 
They are constrained by the $\bar KN$ scattering length, 
in order to see how these inputs constrain the $\pi\Sigma$ threshold behavior 
and the $\Lambda(1405)$ pole positions. 
The results are summarized in Table~\ref{tbl:Martin}.

\begin{table}[bp]
    \centering
    \caption{Summary of numerical results. 
    The model parameters are determined so as to reproduce 
    a value of the $\bar KN$ scattering length with $I=0$, 
    $a_{\bar KN}=-1.70+0.68i$ fm.  
    In each model, pole 1 is found as a resonance
    located between the $\pi\Sigma$ and $\bar KN$ threshold. 
    For pole 2, (R), (B) and (V) denote 
    resonance, bound state and virtual state 
    in the $\pi\Sigma$ channel, respectively.  }
    \begin{tabular}{|l|cccc|}
        \hline
        Model & A1 & A2 & B E-dep & B E-indep  \\
        \hline
        parameter (${\pi\Sigma}$) & $d_{\pi\Sigma}=-1.67$ & $d_{\pi\Sigma}=-2.85$ &
        $\Lambda_{\pi\Sigma}= 1005$ MeV & $\Lambda_{\pi\Sigma}= 1465$ MeV	\\
        parameter (${\bar{K}N}$) & $d_{\bar KN}=-1.79$ & $d_{\bar KN}=-2.05$  &
        $\Lambda_{\bar KN}= 1188$ MeV & $\Lambda_{\bar KN}= 1086$ MeV\\
        \hline
        pole 1 [MeV] & $1422-16i$ & $1425-11 i$ & $1422-22i$ & $1423-29i$ \\
        pole 2 [MeV] & $1375-72 i$ (R) & $1321$ (B) & $1349-54i$ (R) & $1325$ (V) \\
        $a_{\pi\Sigma}$ [fm]  & $0.934$ & $-2.30$ & $1.44$ & $5.50$  \\
        $r_e$ [fm]  & $-5.02$ & $-5.89$ & $-3.96$ & $-0.458$  \\
        $a_{\bar{K}N}$ [fm] (input) & $-1.70+0.68i$ & $-1.70+0.68i$& $-1.70+0.68i$ & $-1.70+0.68i$  \\
        \hline
    \end{tabular}
    \label{tbl:Martin}
\end{table}

First of all, 
we find that the $\bar KN$ scattering length constrains the position of
%%%
the pole near the $\bar KN$ threshold, which we call the (higher) $\Lambda(1405)$
pole,
%%% 
well around $1420 - 20 i$ MeV in all of our models. 
Thus, the value of the $\bar KN$ scattering length with $I=0$ 
can be one of the important quantities to fix the pole position of $\Lambda(1405)$ 
which strongly couples to the $\bar KN$ channel.
On the other hand, 
the $\pi\Sigma$ scattering length and effective range are obtained 
with very different values. 
This is a consequence of different predictions 
of the lower pole positions.
This means that the $\bar KN$ scattering length alone cannot 
constrain the scattering amplitude at far below threshold. 
In contrast, the $\pi\Sigma$ scattering length and effective range are sensitive to 
the lower pole position, 
which will give important constraints on the subthreshold extrapolation 
of the $\bar{K}N$ amplitude. 
In models A1 and B~E-dep we use essentially the same interaction kernel 
but take different off-shell dependence. 
The results are qualitatively the same but there are quantitative differences 
in the $\pi\Sigma$ threshold parameters. 
Thus, we find that the 
additional
%%%
$\pi\Sigma$ threshold quantities 
can be useful also to constrain 
%%%
dynamics of the $\bar KN$-$\pi\Sigma$ coupled channels below the $\bar KN$ threshold.
%%%

In the model A2 and the model B with energy independent potential, 
the $\pi\Sigma$ interaction is so strong that 
they give a bound state and a virtual state, respectively. 
In both cases, the $\pi\Sigma$ attraction is effectively enhanced by the cutoff parameter. 
Though the presence of such a bound or virtual state is 
in contradiction with the result of the 
%%%
more refined 
%%%
calculations 
with the chiral unitary approaches, which take account of the $\bar K^{-}p$ scattering data,
it is important to clarify the position of the lower $\Lambda(1405)$ pole by 
experimental observation.
This can be done by observing the sign and order of magnitude of the $\pi\Sigma$ scattering length. 
If the $\pi\Sigma$ scattering length in the $I=0$ channel would have a negative value, 
there could be a bound state of $\pi$ and $\Sigma$ with $I=0$. 
If the scattering length is positive with as large a value as 5 fm, 
there is a virtual state below 
and close to the $\pi\Sigma$ threshold. 
This also shows the relevance of the $\pi\Sigma$ threshold behavior 
for the subthreshold extrapolation of the $\bar{K}N$ amplitude.

It is also interesting to mention that the present model provides the higher 
$\Lambda(1405)$ pole around $1420 - 20 i$ MeV irrespectively of the off-shell 
behavior and the subthreshold extrapolation to much lower energy region, 
when we use
the Martin's value of the $\bar KN$ scattering length with 
$I=0$ to fix two parameters appearing in the unitarization procedure.
This fact was implicitly shown in Ref.~\citen{Ikeda:2007nz}.
Since, in our simple two-channel model, the interaction is fixed 
by chiral effective theory, we would say that this is a 
consequence of 
the unitarization of the leading order chiral interaction.
Nevertheless, for the definite conclusion on the pole position, 
it is certainly necessary to refine the analysis of Martin done in 80's
by including new measurements of $K^{-}p$, such as
KEK, DEAR and forthcoming SIDDHARTA data, and also 
to use more sophisticated models, for instance, including other channels,
especially $I=1$ amplitudes and isospin breaking effects.   
%%%

\begin{table}[btp]
    \centering
    \caption{Summary of numerical results in the $\pi\Sigma$ single calculations by 
    switching off the $\bar KN$ channel effects.
    The model parameters are same as Table~\ref{tbl:Martin}. 
}
    \begin{tabular}{|l|cccc|}
        \hline
        Model & A1 & A2 & B E-dep & B E-indep  \\
        \hline
        pole 2 [MeV] & $1366-88i$ (R) & $1331$ (V) & $1345-79i$ (R) & $1268$ (V) \\
        $a_{\pi\Sigma}$ [fm]  & $0.77$ & $34.4$ & $0.935$ & $1.42$  \\
        $r_e$ [fm]  & $-4.59$ & $-4.89$ & $-3.58$ & $-0.831$  \\
        \hline
    \end{tabular}
    \label{tbl:single}
\end{table}

%%%
In order to see the $\pi\Sigma$-$\bar KN$ coupled-channels effects,
we calculate the scattering length, the effective range and 
the pole position in the $\pi\Sigma$ single channel by 
switching off the $\bar KN$ channel. We use the same model parameters 
as before, which are determined by the $\bar KN$ scattering length in the
full coupled channel. The results are summarized in Table~\ref{tbl:single},
which shows that
the coupled-channels effects 
contribute moderately as an attractive interaction to the $\pi\Sigma$ channel
at the threshold, which can be seen in the values of the scattering 
length.
%%%
The scattering lengths in the model A1 and B E-dep slightly change,
while in the model A2 and B E-indep the coupled-channels
effects are largely seen in the scattering length. In any cases, 
the pole position is consistent with the classification of the poles 
in terms of $(a,r_{e})$ given in Appendix. 
%%%

\begin{table}[tb]
    \caption{Summary of numerical results with
    the $\Lambda(1405)$ pole position being $1406.5- 25i$ MeV.  
    In each model, pole 1 is found as a resonance
    located between the $\pi\Sigma$ and $\bar KN$ threshold. For pole 2, (R), (B) and 
    (V) denote resonance, bound state and virtual state in the $\pi\Sigma$ channel,
    respectively.}
    \begin{tabular}{|l|cccc|}
        \hline
          Model & A1$^{\prime}$ & A2$^{\prime}$ & B$^{\prime}$ E-dep & B$^{\prime}$ E-indep  \\
      \hline
        parameter (${\pi\Sigma}$) & $d_{\pi\Sigma}=-1.66$ & $d_{\pi\Sigma}=-2.68$ &
        $\Lambda_{\pi\Sigma}= 860$ MeV & $\Lambda_{\pi\Sigma}= 1070$ MeV	\\
        parameter (${\bar{K}N}$) & $d_{\bar KN}= -1.93 $ & $d_{\bar KN}=-2.44$  &
        $\Lambda_{\bar KN}= 1290$ MeV & $\Lambda_{\bar KN}= 1175$ MeV\\
        \hline
        pole 1 [MeV] (input) & $1406-25i$ & $1407-25i$ &  $1405 - i25$ & $1406-i26$ \\
        pole 2 [MeV] & $1381-65i$ (R) & $1304$ (B) &  $1351-i64$ (R) & $1306$ (V) \\
        $a_{\pi\Sigma}$ [fm] & $1.03$ & $-0.75$   &  $1.38$ & $2.77$ \\
        $r_e$ [fm]  & $-5.23$ & $-14.32$ &  $-3.90$ & $-0.509$ \\
        $a_{\bar{K}N}$ [fm] & $-1.37+0.41i$ & $-0.96+0.24i$&  $-1.48+i0.38$ & $-1.67+i0.40$ \\
        \hline
    \end{tabular}
    \label{tbl:Dalitz}
\end{table}

As mentioned above,
%%%
the scattering length $a_{\bar KN}=-1.70 + i0.68$ fm leads to the pole of $\Lambda(1405)$ around 1420 MeV. Although the pole position is not a direct observable, the obtained value deviates from the nominal value shown in PDG~\cite{Nakamura:2010zzi}, $1406.5 - 25 i$ MeV. To examine a different input, we set up the models so as to reproduce the pole position of the PDG value, and perform the same analysis. In this case, we again find two solutions in model A (A1$^{\prime}$ and A2$^{\prime}$)
%%%
and model B (B$^{\prime}$ E-dep and B$^{\prime}$ E-indep).
%%%
Thus, we have four solutions as summarized in Table~\ref{tbl:Dalitz}, which provide different values of the scattering length of the $\bar KN$ channel. This implies that, if $\Lambda(1405)$ is located at as deeply as 30 MeV below the $\bar KN$ threshold, the $\Lambda(1405)$ pole position is also affected by the $\pi\Sigma$ dynamics and the pole position alone cannot constraint the $\bar KN$ threshold quantities. Again, precise information of the $\pi\Sigma$ dynamics is necessary to understand the structure of $\Lambda(1405)$ further and to have models constrained more. It is also important mentioning that, even in such two-body $\pi\Sigma$ and $\bar KN$ dynamics, both of these channels are important and a quasibound state with 30 MeV binding energy is generated by a consequence of both $\pi\Sigma$ and $\bar KN$ dynamics. For investigation of more deeply bound systems, which would be seen in $\bar KNN$, since the $\pi\Sigma$ dynamics should become more important in such systems, one needs experimental information of the $\pi\Sigma$ interaction and more serious consideration of the $\pi\Sigma$ channels.

\begin{table}[tbp]
    \centering
    \caption{Scattering lengths and effective ranges 
    of the $\pi\Sigma$ channel with $I=0$ 
    in chiral unitary approaches
    and the phenomenological potential
    model (EAY).
    The value of $f_{\pi}$, the pole positions of 
    $\Lambda(1405)$ and the result of  
    $\bar K N$ scattering length with $I=0$ in each model 
    are also listed. 
    The calculation is performed with isospin symmetric masses.}
    \begin{tabular}{|c|cccc|c|}
        \hline
        model & ORB~\cite{Oset:2001cn} 
        & HNJH~\cite{Hyodo:2002pk}
        & BNW~\cite{Borasoy:2005ie}
        & BMN~\cite{Borasoy}  
        & EAY~\cite{Esmaili:2009iq}
        \\
        \hline
 	pole 1 [MeV] & $1427 - 17 i$   
 	    & $1428 - 17 i$ 
	    & $1434 - 18 i$ 
	    & $1421 - 20 i$  
	    & $1409 - 20 i$\\
 	pole 2 [MeV] & $1389 - 64 i$ 
	    & $1400 - 76 i$
	    & $1388 - 49 i$ 
	    & $1440 - 76 i$ 
	    & $1299$ (V) \\	 
    $a_{\pi\Sigma}$ [fm] & 0.789 
        & 0.693 
        & 0.770 
        & 0.517  
        & 2.24 \\
    $r_{e}$ [fm] & $-6.27$
        & $-6.50$
        & $-7.18$
        & $-8.63$ 
        & $-0.447$\\
    $a_{\bar KN}$ [fm] & $-1.65 + 0.89 i$ 
        & $-1.63 + 1.01 i$ 
        & $-1.49 + 1.11 i$ 
        & $-1.45 + 0.87 i$ 
        & $-1.76 + 0.42 i$ \\
    $f_{\pi}$ [MeV] & 103.7652 
        & 106.95 
        & 111.2 
        & 120.9  
        &  --- \\
	 \hline
    \end{tabular}
    \label{tbl:ChUM}
\end{table}

It is also interesting to compare the present results with those obtained in more refined models. In table~\ref{tbl:ChUM}, we show the $\pi\Sigma$ scattering length and effective range calculated in various chiral unitary approaches together with the pole positions of $\Lambda(1405)$. 
We adopt the models with the Weinberg-Tomozawa interaction in Refs.~\citen{Oset:2001cn,Hyodo:2002pk,Borasoy:2005ie,Borasoy} where 
%%%
all the channels consisting of the octet baryons and the octet pseudoscalar mesons with $S=-1$ and $I=0$ ($\pi\Sigma$, $\bar KN$, $\eta \Lambda$ and $K\Xi$) are included
%%%
and the model parameters are determined so as to reproduce the total cross sections of the $K^-p$ scattering and the threshold branching ratios of $K^{-}p \to Y\pi$ with $Y=\Lambda$ or $\Sigma$ 
%%%
in the observed
%%%
strong decay of kaonic hydrogen. For comparison with our two-channel model, as in Ref.~\citen{Hyodo:2007jq}, we use the isospin averaged masses 
with the subtraction constants and the pion decay constant being kept fixed to be the original values in these references.
These results are very similar to our models A1 and B E-dep. The results obtained by the chiral unitary models are qualitatively consistent with each other. Nevertheless, in detail, the positions of the lower pole of $\Lambda(1405)$ and the $\pi\Sigma$ scattering lengths are differently predicted. When the pole is closer to the $\pi\Sigma$ threshold (ORB, BNW), the $\pi\Sigma$ scattering length is relatively large. 
It is also worth mentioning that the parameters of BMN were determined by a $\chi^{2}$ fitting using the currently available $\bar KN$ scattering data and Ref.~\citen{Borasoy} found a large uncertainty of the position of pole 2 in their fitting. 
Thus, the quantitative determination of the $\pi\Sigma$ scattering observables 
could be a good guidance to refine the models for the scattering amplitude further and to locate the position of the lower pole of $\Lambda(1405)$. 
We also show, in Table~\ref{tbl:ChUM}, the result obtained by the phenomenological potential model\footnote{In this calculation we have used the isospin averaged masses given in page~\pageref{mass}
and obtained the pole position at $1409-20i$ MeV as shown in Table~\ref{tbl:ChUM}.
It seems that in Ref.~\citen{Esmaili:2009iq} the value of the $\bar KN$ threshold is given by sum of the proton and $K^{-}$ masses. With these values, we have reproduced the $\Lambda(1405)$ pole position at $1405-20i$ MeV.} developed in Ref.~\citen{Esmaili:2009iq}, in which $\pi\Sigma$-$\bar KN$ coupled channels with $I=0$ were considered with an energy-independent potential. The interaction range and  strengths of $\bar KN$-$\bar KN$ and $\pi\Sigma$-$\bar KN$ were determined by $K^{-}p$ scattering data, kaonic atom data and the pole position of $\Lambda(1405)$ at $1405-20i$ MeV~\cite{AY02}, while the interaction strength of $\pi\Sigma$-$\pi\Sigma$ was fixed so that the ratio of the 
whole potential
%%%
strengths of $\pi\Sigma$ and $\bar KN$ is to be 4/3 as motivated by the chiral interaction,
%%%
which is, however, slightly different from the interaction strength in the chiral model given in Eq.~(\ref{eq:WTEindep}). The factor $C_{ij}$ in Eq.~(\ref{eq:WTEindep}) does have the $\pi\Sigma/\bar KN$ ratio equal to 4/3 according to Eq.~(\ref{eq:Cij}). (These two potential strengths coincide in the flavor SU(3)  limit.) The result of the phenomenological model is found to be
%%%
very similar to that obtained in B$^{\prime}$ E-indep given in Table~\ref{tbl:Dalitz}. This is because the phenomenological potential model uses the energy-independent interaction and the parameters were fixed by the $\Lambda(1405)$ pole position at $1405 - 20i$ MeV. In addition, the $\pi \Sigma$ interaction strength is similar to the chiral models.
%%%
%\footnote{To be precise, the $\pi\Sigma$-$\pi\Sigma$ interaction strength in the EAY potential model was fixed so as to make the $\pi\Sigma/\bar KN$ ratio of the whole potential strengths to be 4/3. This is not exactly the same as the interaction strength in the chiral model even though one takes the static approximation for the Weinberg-Tomozawa interaction as in the B$^{\prime}$ E-indep.\ model given Eq.~(\ref{eq:WTEindep}), in which the factor $C_{ij}$ does have the $\pi\Sigma/\bar KN$ ratio equal to 4/3 according to Eq.~(\ref{eq:Cij}). (These two potential strengths coincide in the flavor SU(3)  limit.)  In addition, different form factors are used in the EAY and the B$^{\prime}$ E-indep, which introduces different off-shell dependence. Even though there are such differences in the two models, it is interesting that they provide very similar results.} 
%%%
These are also the case in B$^{\prime}$ E-indep. A virtual state is found also in the phenomenological potential model due to the attractive interaction in the $\pi\Sigma$ channel. 
%%%

The $\pi\Sigma$ channel is the lowest energy meson-baryon state with $I=0$ and $S=-1$, and it is believed that there is no bound state below the threshold. One may consider that such a lowest state with one pion may be well described by the chiral perturbation theory, as in the case of the $\pi N$ system. Using the Weinberg-Tomozawa interaction (\ref{eq:WTterm}) with $C_{\pi\Sigma}=4$, we obtain the scattering length and effective range at the leading order of the chiral perturbation theory as
\begin{eqnarray}
  a_{\pi\Sigma} &=& 
  \frac{\mu_{\pi\Sigma}}{2\pi f_{\pi}^{2}} = 0.455 \text{ fm}, \\
  r_{e}  &=& -\frac{\pi}{\mu_{\pi\Sigma}} 
  \frac{f_{\pi}^{2}(m_{\pi}^{2}+2M_{\Sigma}^{2})}{m_{\pi}^{2}M_{\Sigma}^{2}}
  = -4.52 \text{ fm},
\end{eqnarray}
with the $\pi\Sigma$ reduced mass $\mu_{\pi\Sigma}=m_{\pi} M_{\Sigma}/(m_{\pi}+M_{\Sigma}) $. Comparing the results obtained by the chiral perturbation theory with those calculated in dynamical models, we find that the deviation of the results of the dynamical models from those of the chiral perturbation theory is much larger than 20\%, in contrast to the $\pi N$ scattering case where the perturbative calculation works well. This is understood by the existence of the pole singularity around the threshold. The interaction strength in the $\pi\Sigma$ channel is $C_{\pi \Sigma}=4$, which is close to the critical coupling constant to have a bound state~\cite{Hyodo:2006yk}. The strong attraction generates a pole around the threshold and hence the perturbative calculation may not be applicable. On the other hand, the coupling constant in the $\pi N$ channel is $C_{\pi N}=2$ ($C_{\pi N}=-1$) for $I=1/2$ ($I=3/2$), which is weaker than the $\pi\Sigma$ case (repulsive) so that no pole singularity appears around the threshold. In this way, the determination of the $\pi \Sigma$ threshold quantities with $I=0$ gives us an interesting test for the applicability of the chiral perturbation theory. 

%%%%%%%%%%%%%%%%%%%%%%%%%%%%%%%%%%%%%%%%%%%%%%%%%%%%%%%%%%%%%%%%%%
\section{Conclusion}
%%%%%%%%%%%%%%%%%%%%%%%%%%%%%%%%%%%%%%%%%%%%%%%%%%%%%%%%%%%%%%%%%%

We have investigated the $\pi\Sigma$ scattering length and effective range in various models, solving Lippmann-Schwinger equation for the $\pi\Sigma$ and $\bar KN$ coupled channels with the interaction kernel obtained by the leading order chiral perturbation theory. We have used different regularization schemes to integrate the Lippmann-Schwinger equation, which is equivalent to applying different treatments of off-shell scattering amplitudes. We find that the constraints from the $\bar KN$ scattering length determine the theoretical descriptions of the scattering amplitude around $\bar{K}N$ threshold and the higher pole position of $\Lambda(1405)$, whereas it is not enough to fix the precise position of the lower energy pole and the scattering amplitude around the $\pi\Sigma$ threshold. Consequently, the $\pi\Sigma$ scattering length and the effective range will give us new additional information for the $\pi\Sigma$-$\bar KN$ dynamics and the structure of $\Lambda(1405)$. 

At this moment, there is no experimental information of the $\pi\Sigma$ threshold quantities. Although scattering experiments are impossible, we may utilize the interference of the final state interaction in the decay of the heavy particle~\cite{Lambdacdecay}, in analogy with the determination of the $\pi\pi$ scattering length in Ref.~\citen{Cabibbo:2004gq}. Moreover, recent development of lattice QCD may bring the information of the scattering length~\cite{Ikedalattice}. As a first step, the sign of the scattering length can be used to check the existence of the bound state below the threshold. In the long run, the determination of the magnitude of the scattering length and the effective range will provide the detailed information of the pole structure of $\Lambda(1405)$ and the severe constraints for the $\pi\Sigma$-$\bar KN$ scattering amplitude. We hope that the present semi-qualitative analysis on the $\pi\Sigma$ scattering length and effective range gives a good guideline for forthcoming experiments and lattice QCD calculations.

\section*{Acknowledgements}
%We would like to thank ...........
Y.I. acknowledges the support by the Japan Society for the Promotion of Science,
Grant-in-Aid for Scientific Research on Innovative Areas (No. 2004:
20105001, 20105003).
T.H.\ thanks the support from the Global Center of Excellence Program by MEXT, 
Japan through the Nanoscience and Quantum Physics Project of the Tokyo Institute 
of Technology. 
This work was partly supported by the Grant-in-Aid for Scientific Research from 
MEXT and JSPS (Nos.\
  22740161, 22105507, % Jido
  21840026, % Hyodo,
  20540270 % Sato
), 
the Grant-in-Aid for the Global COE Program ``The Next Generation of Physics, Spun 
from Universality and Emergence'' from MEXT of Japan,
and 
U.S. Department of Energy, Office of Nuclear 
Physics Division, under Contract No. DE-AC05-06OR23177,
under which Jefferson Science Associates operates the Jefferson Lab.
This work was done in part under the Yukawa International Program for Quark-hadron Sciences (YIPQS).

\appendix

%%%%%%%%%%%%%%%%%%%%%%%%%%%%%%%%%%%%%%%%%%%%%%%%%%%%%%%%%%%%%%%%%%
\section{Scattering length and effective range}\label{sec:length}
%%%%%%%%%%%%%%%%%%%%%%%%%%%%%%%%%%%%%%%%%%%%%%%%%%%%%%%%%%%%%%%%%%

Let us consider an $s$-wave single-channel two-body scattering amplitude $f(k)$ with momentum $k$. This is given in terms of the $S$ matrix element with $l=0$ by 
\begin{equation}
   f(k) = \frac{s_{\ell=0}(k) - 1}{2ki}=\frac{1}{k\cot\delta-ki}\ ,
\end{equation}
where $\delta$ is the $s$-wave phase shift. The scattering length $a$ and the effective range $r_e$ are defined as the expansion coefficients  of $k\cot\delta$ in terms of $k^{2}$ around $k=0$:\footnote{The sign of $a$ 
is opposite to the standard convention in nuclear physics. In our convention, $f(k)\to a$ when $k\to 0$.}
\begin{equation}
    k\cot \delta = \frac{1}{a}+ r_e\frac{k^2}{2}
    +\cdots
    \label{eq:range}
    .
\end{equation}
Taking only the first two terms in Eq.(\ref{eq:range}), we write the scattering amplitude as
\begin{align}
    f(k)=&
    \left(\frac{1}{a}-ki +\frac{r_e}{2}k^2\right)^{-1} .
    \nonumber 
\end{align}
This amplitude has a pole, when the momentum $k$ satisfies the condition
\begin{align}
    \frac{r_e}{2}k^2 - ki +\frac{1}{a}
    =&0  \ .
    \label{eq:pole}
\end{align}

If $r_e=0$, the solution of Eq.~\eqref{eq:pole} is given by
\begin{align}
    k
    =&-\frac{i}{a} \ .
    \label{eq:solution_a} 
\end{align}
If there are no open channels below the threshold ($k=0$), the scattering amplitude is real at $k=0$, and consequently the scattering length $a$ is also real. In this case, the pole singularity given by Eq.~(\ref{eq:solution_a}) lies on the imaginary axis on the complex $k$ plane. The solution with $\im k>0$ ($\im k<0$) corresponds to the bound (virtual) state, so we have a bound state for $a<0$ and a virtual state for $a>0$. When the bound state is formed, the scattering length looks as if the interaction is repulsive. 

\begin{figure}[tbp]
    \centering
    \includegraphics[width=0.4\textwidth,clip]{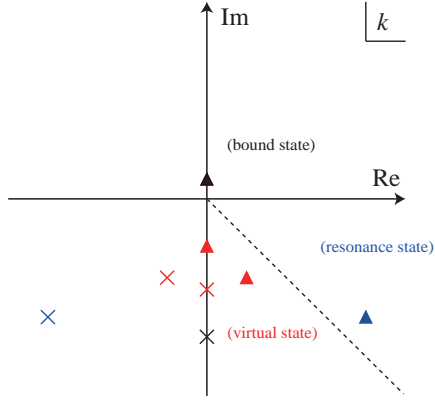}
    \caption{\label{fig:pole}
    Schematic illustration of the pole position and the 
    resonance/virtual/bound state
    in the complex $k$ plane.
    Triangles represent the poles which are relevant 
    to the amplitude above the threshold,
    and the crosses denote the other poles. 
    The dotted line ($\re E_{\text{pole}}=0$) 
    is the boundary between the regions of resonances 
    and virtual states. 
    }
\end{figure}

Next, we study the case of $r_e\neq 0$. In this situation, the inverse of the scattering amplitude is given by a quadratic function, so that we always have a pair of poles for given $a$ and $r_e$ except for the case of two-fold root. The solutions of Eq.~\eqref{eq:pole} are
\begin{align}
    k
    =&  \frac{i}{r_e}
    \pm \frac{1}{r_e}\sqrt{ -\frac{2r_e}{a} -1 } .
    \label{eq:solution}
\end{align}
Defining $-2r_e/a-1\equiv D$, the solutions can be classified by the sign of $D$:
\begin{itemize}
    \item  When $D<0$, the solutions are pure imaginary. 

    \item  When $D>0$, the solutions are complex with the same imaginary part. 

\end{itemize}
The case $D<0$ corresponds to 
a bound state or a virtual state without width,
while 
a virtual state with finite width or a resonance
can be formed for the case $D>0$. For later convenience, we schematically illustrate the pole position and the resonance/virtual/bound state in Fig.~\ref{fig:pole}. 
The poles with $\im k>0$ ($\im k<0$) are mapped onto the first (second) Riemann sheet of the energy plane. As mentioned, there are always two poles for given $(a,r_e)$, but the scattering amplitude above the threshold is mainly affected by the pole close to 
the scattering region $\re k \ge 0$
(triangles in Fig.~\ref{fig:pole}). In the following, we classify the solutions by the character of the pole most relevant
to the amplitude, 
which appears at larger $\im k$ for the state without the width or at $\re k >0$ for the state with the width.
%%%

%--figure---------------------------------
%--figure---------------------------------

Let us first consider the $D<0$ case. The parameter region for $D<0$ is given by
\begin{equation}
    \begin{cases}
        a>0, \quad r_e > -a/2  \\
        a<0, \quad r_e < -a/2  
    \end{cases} .
    \label{eq:Dneg1}
\end{equation}
The pole positions are
\begin{align}
    \re k_{\text{pole}}
    =& 0 , \quad
    \im k_{\text{pole}}
    = \frac{1}{r_e}\left(+1\pm \sqrt{-D}\right) .
    \nonumber
\end{align}
In this case, both the two solutions lie on the imaginary $k$ axis. We can further classify the solutions by the values of  $r_e$ and $D$:
\begin{equation}
    \begin{cases} 
	r_e>0,\quad \sqrt{-D}>1, & \text{virtual state without width} \\
    r_e>0,\quad \sqrt{-D}<1,
	& \text{bound state } \\
    r_e<0,\quad \sqrt{-D}>1, 
    & \text{bound state } \\
    r_e<0,\quad \sqrt{-D}<1, & 
    \text{virtual state without width}
    \end{cases}
    \label{eq:Dneg2}
\end{equation}
Note that $\sqrt{-D}>1$ means $2r_e/a >0$, which corresponds to the case where $a$ and $r_e$ has the opposite sign. Thus, in terms of $a$ and $r_e$, we have
\begin{equation}
    \begin{cases} 
        a>0,\quad r_e >-a/2 ,
        & \text{virtual state without width} \\
        a<0, \quad r_e < -a/2  ,
        & \text{bound state}
    \end{cases}
    \label{eq:Dneg3} 
\end{equation}

\begin{figure}[htbp]
    \centering
    \includegraphics[width=0.6\textwidth,clip]{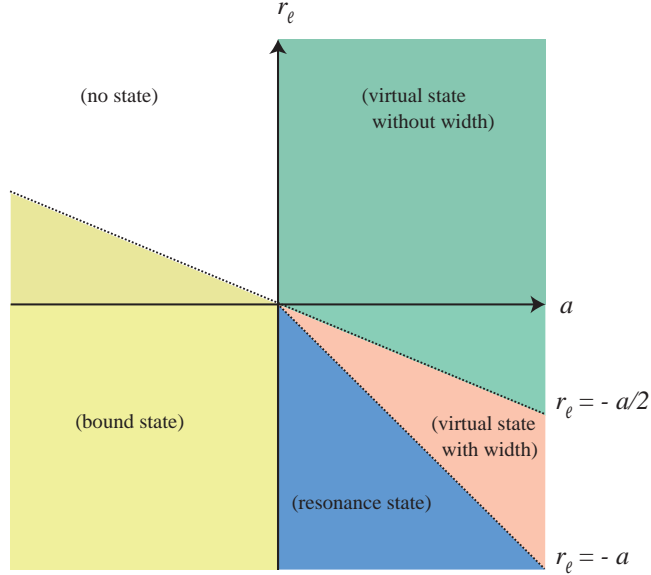}
    \caption{\label{fig:arplot}
    Parameter region corresponds to resonance/virtual/bound state 
    in $a$-$r_e$ plot.}
\end{figure}

Next we consider the $D>0$ case. The parameter region is given by
\begin{equation}
    \begin{cases}
        a>0, \quad r_e < -a/2  \\
        a<0, \quad r_e > -a/2  
    \end{cases} .
    \label{eq:Dpos1}
\end{equation}
The real and imaginary parts of the pole can be extracted 
from Eq.~\eqref{eq:solution} as
\begin{align}
    \re k_{\text{pole}}
    =&\pm \frac{1}{r_e}\sqrt{D}, \quad
    \im k_{\text{pole}}
    = \frac{1}{r_e} .
    \nonumber
\end{align}
In this case, the pair of poles should emerge in the symmetric position with respect to the imaginary $k$ axis. If $r_e >0$, two poles with finite width appear in the first Riemann sheet of the complex energy plane. 
%%%
Since such poles are forbidden by the causality, we consider them to be 
an artifact of the truncated effective range expansion and expect them to disappear 
when higher order terms in $k$ are included. We therefore classify the region 
of $(a,r_e)$ giving rise to such poles as ``no state''.
From Eq.~\eqref{eq:Dpos1}, we find that this is the case for $a<0$.
%%%
For $r_e <0 $, the two poles
are interpreted as a virtual state with width or a resonance state, depending on the ratio of the real and imaginary parts. The boundary in the complex $k$ plane for the virtual and resonance state (with nonrelativistic kinematics $E_{\text{pole}}=k_{\text{pole}}^2/2\mu$) is given by $\re E_{\text{pole}}=0$ in the second Riemann sheet, namely
\begin{equation}
    \begin{cases}
        \im k_{\text{pole}}<0,\quad |\re k_{\text{pole}}|/|\im k_{\text{pole}}|
	>1, & \text{resonance state}  \\
        \im k_{\text{pole}}<0,\quad |\re k_{\text{pole}}|/|\im k_{\text{pole}}|
	<1, & \text{virtual state with width}
    \end{cases}
    \label{eq:cases}
\end{equation}
The classification of the solution is then given by
\begin{equation}
    \begin{cases}
        \sqrt{D}>1, & \text{resonance state}  \\
        \sqrt{D}<1, & \text{virtual state with width}
    \end{cases}
    \label{eq:Dpos2}
\end{equation}
Thus, combining \eqref{eq:Dpos1} and \eqref{eq:Dpos2}, the parameter space for obtaining resonance/virtual state is given by
\begin{equation}
    \begin{cases}
	a>0, \quad
	r_e <-a,
	& \text{resonance state}  \\
	a>0, \quad
	-a/2 > r_e >- a, & \text{virtual state with width} \\
	a<0, \quad r_e >- a/2, 
	& \text{no state}
    \end{cases}
    \label{eq:Dpos3}
\end{equation}
We summarize Eq.~\eqref{eq:Dneg3} and \eqref{eq:Dpos3} in $a$-$r_e$ plot in Fig.~\ref{fig:arplot}. If $a$ is negative, there is a bound state. For positive $a$, the ratio of $a$ and $r_e$ determines the location of the singularities. Setting $r_e=0$, we recover the classification by the scattering length Eq.~\eqref{eq:solution_a}.

%--figure---------------------------------
%--figure---------------------------------

To illustrate the relation between the threshold parameters and pole position of the amplitude, we solve the single-channel version of the model shown in section~\ref{sec:modelA}. We choose the $\pi\Sigma$ channel where we obtain a resonance state with the subtraction constant $d=-2$. The scattering length and effective range are
\begin{equation}
    a=1.06 \text{ fm}, \quad r_{e} = -4.67 \text{ fm}   ,
    \nonumber 
\end{equation}
in accordance with the resonance solution in~\eqref{eq:Dpos3}, $r_e<-a<0$. When we increase the absolute value of the subtraction constant, the interaction strength is effectively enhanced~\cite{Hyodo:2008xr}, so the pole is expected to become a virtual state and eventually a bound state. 

\begin{figure}[tbp]
\centering
\includegraphics[width=0.45\textwidth,clip]{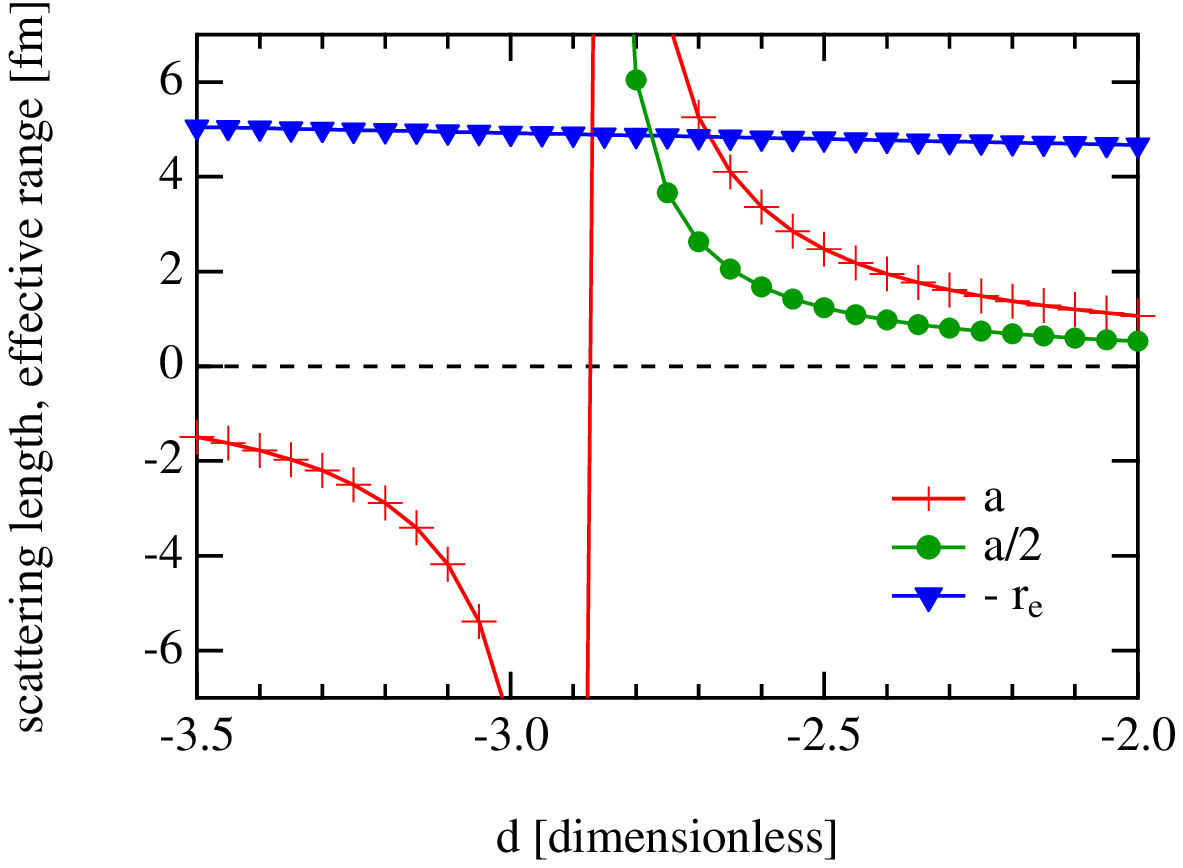}
\includegraphics[width=0.45\textwidth,clip]{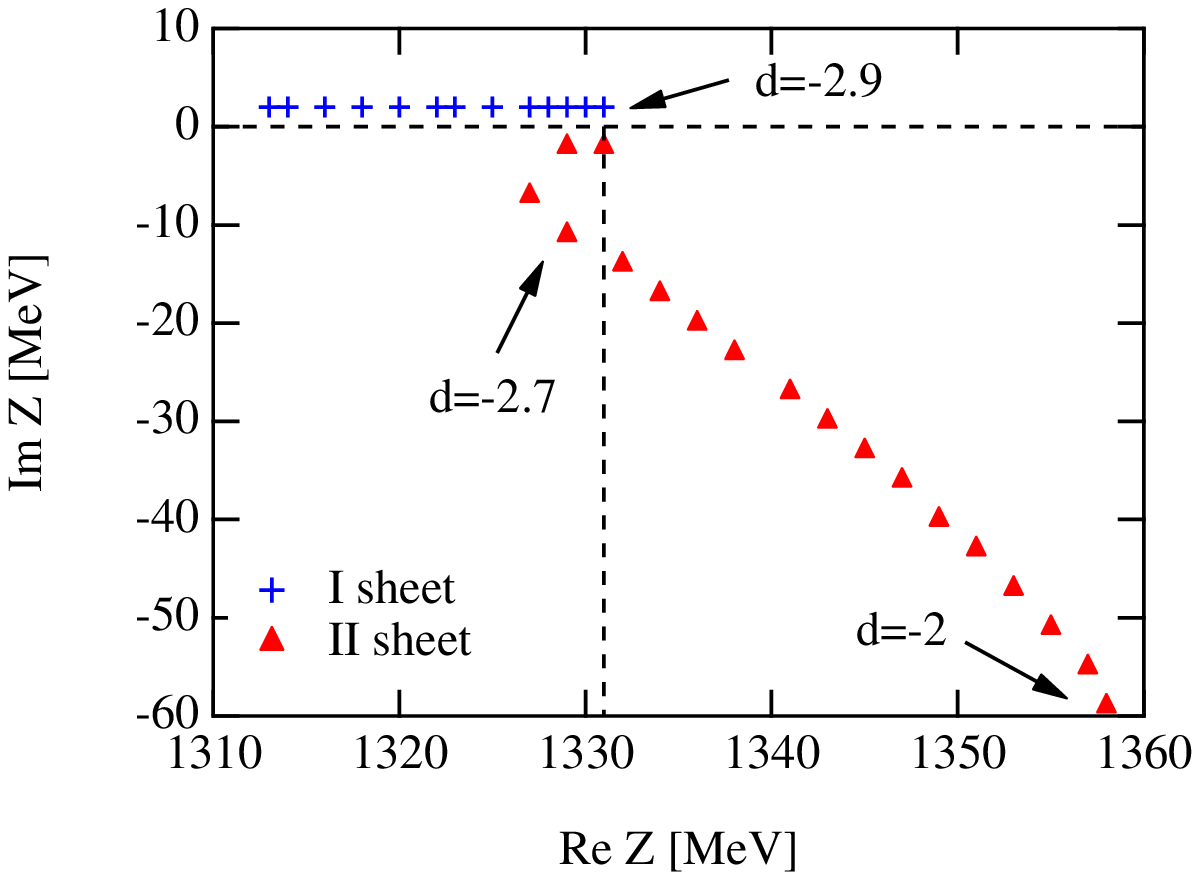}
\caption{The $\pi\Sigma$ threshold quantities (left) and the pole position of the scattering amplitude (right) with various subtraction constants $d$. All symbols are plotted at each 0.5 step of the subtraction constant. In the left panel, the effective range is plotted with opposite sign and $a/2$ is also shown for positive scattering length, for convenience of the comparison with Eq.~\eqref{eq:Dpos3}. In the right panel, there are always two poles in the complex energy plane, but we only show the most relevant pole to the observables defined on the real axis above threshold. The poles on the first (second) Riemann sheet are plotted by the crosses (triangles), which are slightly shifted to positive (negative) direction of the imaginary axis when they are on the real axis.}
\label{fig:are}
\end{figure}

We calculate the scattering length, the effective range and the pole positions by varying the subtraction constant $d$ from $-2$ to $-3.5$ as shown in Fig.~\ref{fig:are}. The effective range is shown with opposite sign for comparison. As the absolute value of the subtraction constant is increased, both the scattering length and effective range $(-r_e)$ become large. The resonance pole moves to the lower energy region with decreasing the width. This indicates the effective enhancement of the interaction strength.

%--figure---------------------------------
%--figure---------------------------------

In contrast to the mild $d$ dependence of the effective range, the scattering length increase rapidly for $d\lesssim -2.6$. At $d=-2.7$, the scattering length becomes larger than the effective range with negative sign. At the same time, the resonance pole moves below the threshold and becomes a virtual state with finite width. When the subtraction constant reaches $d=-2.8$, the half of the scattering length exceeds the effective range $a/2>-r_e$. This coincides with the appearance of the virtual state pole on the real axis. As we further increase the magnitude of the subtraction constant, the pole finally becomes a bound state for $d\lesssim-2.9$. Before that, the scattering length becomes very large positive value. Once the bound state is formed, the scattering length is obtained as a negative value whose absolute value is decreasing. These behaviors are fairly consistent with the classification given in Eqs.~\eqref{eq:Dneg3} and \eqref{eq:Dpos3}.

\end{document}